\documentclass[aps,nofootinbib]{revtex4}
\usepackage{amsmath}
\usepackage{graphicx}

\begin{document}

\title{Deconstructing triplet nucleon-nucleon scattering}
\author{Michael C. Birse}
\affiliation{Theoretical Physics Group, School of Physics and Astronomy\\
The University of Manchester, Manchester, M13 9PL, UK\\}

\begin{abstract}

Nucleon-nucleon scattering in spin-triplet channels is analysed 
within an effective field theory where one-pion exchange is treated 
nonperturbatively. Justifying this requires the identification of an 
additional low-energy scale in the strength of that potential. Short-range 
interactions are organised according to the resulting power counting, 
in which the leading term is promoted to significantly lower order than 
in the usual perturbative counting. In each channel there is a critical 
momentum above which the waves probe the singular core of the tensor 
potential and the new counting is necessary. When the effects of 
one- and two-pion exchange have been removed using a distorted-wave Born 
approximation, the residual scattering in waves with $L\leq 2$ is well 
described by the first three terms in the new counting. In contrast, the 
scattering in waves with $L\geq 3$ is consistent with the perturbative 
counting, at least for energies up to 300~MeV. This pattern is in 
agreement with estimates of the critical momenta in these channels.

\end{abstract}
\maketitle

\vskip 10pt

\section{Introduction}

Over the last fifteen years, considerable effort has gone into trying
to analyse nuclear forces using the systematic tools of effective field
theory (EFT).\footnote{For reviews, from various points of view,
see Refs.~\cite{border,bvkrev,eprev}.} The starting point was Weinberg's 
original proposal \cite{wein} that these forces could be described within 
the framework of chiral perturbation theory (ChPT). This approach 
organises the terms in the effective Lagrangian or Hamiltonian 
according to powers of low-energy scales they contain. These scales,
generically denoted $Q$, include momenta and the pion mass. In this 
``Weinberg" power counting, the leading terms of the nucleon-nucleon
potential are one-pion exchange (OPE) and an energy-independent contact 
interaction, both of which are of order $Q^0$. 

Weinberg also noted the enhancement of the nonrelativistic two-nucleon
propagator near threshold and proposed that the leading terms in the 
potential should be iterated to all orders in order to generate 
nonperturbative effects, such as the deuteron bound state. This approach, 
referred to here as the ``Weinberg--van-Kolck" (WvK) scheme, has been 
widely applied by van Kolck and collaborators and by many 
others.\footnote{Examples of successful applications can be found the 
reviews cited in footnote 1.} However, even with this enhancement, 
nucleon-nucleon loop integrals are of order $Q$. Although this is lower 
than the order $Q^2$ expected in a relativistic theory, it means that 
each iteraction of the leading potential in a scattering equation raises 
the order by one power of $Q$. Hence the resulting amplitude should still 
be perturbatively expandable in powers of the scales $Q$. 

In order to justify treating the leading terms in the potential 
nonperturbatively, a further IR enhancement is needed, to promote them
to order $Q^{-1}$. Such a promotion is only possible within a consistent 
power counting if we can identify additional low-energy scales in the 
nucleon-nucleon system. In the case of $S$-wave scattering, the large 
scattering lengths provide such scales, and these lead to an EFT 
in which the leading, energy-independent contact terms are treated 
nonperturbatively \cite{bvk,vk,ksw}. At low energies, where the 
finite-range of OPE is not resolved, the resulting expansion of the
potential is simply the effective-range expansion \cite{bethe,newton}.

Although a similar systematic justification for the iteration of OPE was 
not provided, the WvK scheme has been successfully used to describe a 
variety of few-nucleon systems and their interactions. Nonetheless,
its validity in the $^3S_1$--$^3D_1$ channel has been questioned
\cite{bbsvk} and, more recently, several groups have observed that 
Weinberg power counting can break down for nucleon-nucleon scattering in 
spin-triplet channels with nonzero orbital angular momentum 
\cite{ntvk,birse,em}.\footnote{Closely related observations can be 
found in the work of Pav\'on Valderrama and Ruiz Arriola \cite{pvra}.}
In particular, Nogga, Timmermans and van Kolck \cite{ntvk} find that the 
leading contact interactions can be substantially promoted in channels 
where tensor OPE is attractive, although this conclusion does depend on 
the choice of cut-off, as stressed by Epelbaum and Meissner \cite{em}.

To establish a quantitative form for this new power counting, we
need first to identify a low-energy scale that would justify iterating OPE,
and then to analyse the scale dependence of the associated short-range
interactions. The renormalisation group (RG) \cite{wrg} provides the 
natural tool for such an analysis. In its Wilsonian version, it has been 
applied to two-body scattering by short-range forces, showing that the 
effective-range expansion is based on a nontrivial fixed point of the 
RG flow \cite{bmr}. Distorted-wave methods have been used to extend the
approach to systems with known long-range forces \cite{bb1,bb2}.

Once a factor of $1/M_{\scriptscriptstyle N}$ has been divided out of the
Hamiltonian, the strength of the OPE potential can be expressed in terms
of the momentum scale
\begin{equation}
\lambda_\pi=\frac{m_\pi^2}{f_{\pi{\scriptscriptstyle NN}}^2
M_{\scriptscriptstyle N}}\simeq 290\;\mbox{MeV}.
\label{eq:lambdapi}
\end{equation}
where $f_{\pi{\scriptscriptstyle NN}}$ is the pseudovector $\pi$N coupling 
constant. In the chiral limit, it can be written
\begin{equation}
\lambda_\pi=\frac{16\pi F_\pi^2}{g_{\scriptscriptstyle A}^2 
M_{\scriptscriptstyle N}},
\end{equation}
where $M_{\scriptscriptstyle N}$ the nucleon mass, $g_A$ is the axial
coupling of the nucleon and $F_\pi$ the pion decay constant. In strict
ChPT, $\lambda_\pi$ is therefore a high-energy scale, built out of 
$4\pi F_\pi$ and $M_{\scriptscriptstyle N}$. None-the-less its numerical value 
is small, only about twice $m_\pi$. As a result, perturbative treatments of 
OPE (as advocated by Kaplan, Savage and Wise \cite{ksw}) fail to converge or
converge only slowly \cite{fms,bbsvk}. This suggests that we should explore 
the consequences of identifying $\lambda_\pi$ as a low-energy scale, counting 
it as of order $Q$. Since $\lambda_\pi$ is proportional to 
$1/M_{\scriptscriptstyle N}$, this can be thought of as a concrete version of
Weinberg's suggestion that $1/M_{\scriptscriptstyle N}$ should be treated as 
if it were of order $Q$ \cite{wein}.

In Ref.~\cite{birse}, I applied a renormalisation group analysis to the
short-range potential in the presence of tensor OPE. This made use of the 
distorted waves (DW's) of a $1/r^3$ potential (the chiral limit of tensor 
OPE). In the resulting power counting, the leading short-range potential is 
of order $Q^{-1/2}$, independently of the orbital angular momentum. This is
quite different from Weinberg counting, where the leading term in the
$L$-th partial wave is of order $Q^{2L}$. Subleading terms containing
powers of the energy appear at orders $Q^{3/2}$, $Q^{7/2}$, and so on. This 
promotion of short-range terms confirms the numerical observations of Nogga, 
Timmermans and van Kolck \cite{ntvk} and makes quantitative the new counting 
proposed there. The terms in the resulting potential can be directly 
related to a DW Born expansion, similar to that in Refs.~\cite{bb1,bb2}.

The validity of this counting does depend on the energies considered, since 
in each channel there is critical momentum above which waves penetrate 
the centrifugal barrier and reach the region where the $1/r^3$ singularity 
dominates. The analyses of Refs.~\cite{ntvk,birse} show that a nonperturbative
treatment of OPE, and hence the new counting, is needed in the $S$, $P$ 
and $D$ waves for momenta of order $m_\pi$. In contrast, waves with 
$L\geq 3$ do not probe the singularity until momenta of $\sim 2$~GeV are
reached. In these higher partial waves OPE can be treated as a perturbation 
and short-distance interactions can be organised according to the usual 
Weinberg power counting.

The results of the analysis of Ref.~\cite{birse} were purely formal, 
leading to the power counting that governs the importance of the terms in the 
expansion of the short-range potential. In the present paper, I explore 
its practical consequences by analysing nucleon-nucleon scattering in 
spin-triplet channels, with an extension of the method applied to 
singlet channels in Ref.~\cite{bmcg}. For simplicity I consider only
the uncoupled waves: $^3P_{0,1}$, $^3D_2$, $^3F_3$ and $^3G_4$. The 
extension of the method to coupled waves such as $^3S_1$--$^3D_1$ or
$^3P_2$--$^3F_2$ is very similar in principle, but is technically more 
complicated because of the matrix nature of the equations.

The RG analysis relies on the forms of the DW's at small radii, where they 
tend to asymptotic forms that are independent of energy. These waves are 
obtained by solving the Sch\"odinger equation, as described in Sec.~II. At 
small enough radii they show nonperturbative behaviour controlled by the 
$1/r^3$ singularity of the tensor potential. In the case of waves with 
$L\leq 2$, this region extends out to about 1~fm. For lab kinetic energies 
up to 300~MeV the waves reach their asymptotic forms only for radii less 
than about 0.6~fm, and there they are dominated by the $1/r^3$ potential. 
This nonperturbative behaviour is present in waves with $L\geq 3$, but only
for only for radii less than about 0.2 fm. In the range 0.2--0.6~fm they 
have the normal power-law forms associated with the centrifugal barrier. This 
confirms the expectations in Refs.~\cite{ntvk,birse} that low partial waves
need the new power counting for energies in this range, whereas the higher 
waves can still be described perturbatively using Weinberg counting.

I then use DW methods to ``deconstruct" empirical scattering amplitudes
by removing the effects of known long-range-forces. The residual
amplitude can then be interpreted directly in terms of an effective 
short-range potential. This technique can provide a better indication 
of how well the known forces are able to describe the scattering, compared 
to simply plotting phase shifts. Such plots can be misleading since they 
tend to hide small differences in peripheral waves at low energies, 
which is just where the long-range forces should dominate. If the resulting 
potential still shows strongly nonlinear energy dependence at low energies, 
then this implies that long-range forces are still making important
contributions. Short-range forces lead to a smooth energy dependence
that can be expanded as a power series.

As in the similar treatment of singlet channels \cite{bmcg}, I take
several Nijmegen PWA's or potentials \cite{nijnn}, to give an indication 
of the uncertainties involved in these analyses of the data. In Sec.~III, 
I use them to construct scattering amplitudes between the DW's of the 
OPE potential and from these I extract short-range potentials in the 
uncoupled spin-triplet channels. The resulting potentials show rapid
energy dependence at low energies, indicating that important long-range 
physics is still present.

The most obvious long-range forces that need to be removed next are 
two-pion exchange (TPE) and relativistic corrections to OPE. These appear 
at orders $Q^2$ and $Q^3$ (in Weinberg counting). I use here the forms
of the TPE potentials given in Refs.~\cite{kbw,nij99} and the corresponding
order-$Q^2$ correction to OPE \cite{friar}. At this order, there is
also a $\gamma\pi$-exchange potential, calculated in Ref.~\cite{fvkpc}.
These potentials can all be subtracted perturbatively using the DWBA. 
The residual short-distance interactions shown in 
Sec.~IV are consistent with the new power counting in the $^3P_{0,1}$ and 
$^3D_2$ waves. In higher waves, $^3F_3$ and $^3G_4$, the uncertainties in 
the Nijmegen PWA's make it hard to draw very strong conclusions but the
residual short-range potentials are smaller after removal of TPE, and 
similar to those in the singlet channels \cite{bmcg}. 

\section{Distorted waves}

The radial Schr\"odinger equation that describes the relative motion of two
nucleons interacting through the long-range OPE potential is
\begin{equation}
-\,\frac{1}{M_{\scriptscriptstyle N}}\left[\frac{{\rm d}^2}{{\rm d}r^2}
+\frac{2}{r}\,\frac{{\rm d}}{{\rm d}r}-\frac{L(L+1)}{r^2}\right]\psi(r)
+\Bigl[V_{\pi {\scriptscriptstyle C}}(r)
+V_{\pi {\scriptscriptstyle T}}(r)\Bigr]
\,\psi(r)=\frac{p^2}{M_{\scriptscriptstyle N}}\psi(r),
\label{eq:se}
\end{equation}
where the central piece of the lowest-order potential is
\begin{equation}
V_{\pi {\scriptscriptstyle C}}(r)=\frac{1}{3}\,f_{\pi{\scriptscriptstyle NN}}^2
\frac{e^{-m_\pi r}}{r}\, ({\boldsymbol\sigma}_1\cdot{\boldsymbol\sigma}_2)
({\boldsymbol\tau}_1\cdot{\boldsymbol\tau}_2),
\end{equation}
and its tensor piece is
\begin{equation}
V_{\pi {\scriptscriptstyle T}}(r)=\frac{1}{3}\,
\frac{f_{\pi{\scriptscriptstyle NN}}^2}{m_\pi^2}
\left(3+3m_\pi r+m_\pi^2 r^2\right)
\frac{e^{-m_\pi r}}{r^3}\, S_{12} ({\boldsymbol\tau}_1
\cdot{\boldsymbol\tau}_2).
\end{equation}
Here the tensor operator, $S_{12}=3({\boldsymbol\sigma}_1\cdot\hat{\bf r})
({\boldsymbol\sigma}_2\cdot\hat{\bf r})
-{\boldsymbol\sigma}_1\cdot{\boldsymbol\sigma}_2$, takes the value
$+2$ in the uncoupled $^3P_1$, $^3D_2$, \dots~channels, and $-4$ in the 
$^3P_0$ channel. The isospin factor, 
${\boldsymbol\tau}_1\cdot{\boldsymbol\tau}_2$, is $+1$ for channels with 
odd $L$ and $-3$ for even $L$.
The on-shell momentum in the centre-of-mass frame, denoted 
by $p$, is related to the lab kinetic energy, $T$, by 
$T=2p^2/M_{\scriptscriptstyle N}$. 

At small enough radii, all the solutions in a given channel tend to a common, 
energy-independent form which is determined by the $1/r^3$ term of the tensor 
potential and the $1/r^2$ centrifugal barrier. This form can be found by 
solving Eq.~(\ref{eq:se}) at zero energy in the chiral limit ($m_\pi=0$) where
it can be written
\begin{equation}
\left[\frac{{\rm d}^2}{{\rm d}r^2}
+\frac{2}{r}\,\frac{{\rm d}}{{\rm d}r}-\frac{L(L+1)}{r^2}
-\frac{\beta_{LJ}}{r^3}\right]\psi_0(r)=0,
\label{eq:zese}
\end{equation}
Here I have specialised to the case of uncoupled triplet channels and I have 
introduced the length scale
\begin{equation}
\beta_{LJ}=\left\{\begin{array}{rl} -4/\lambda_\pi, &\quad L=1,\ J=0,\cr
\noalign{\vspace{5pt}}
+2/\lambda_\pi, &\quad L=J\ \mbox{odd,}\cr
\noalign{\vspace{5pt}}
-6/\lambda_\pi, &\quad L=J\ \mbox{even.}\end{array}\right. 
\end{equation}
The solutions in this limit can be expressed in terms of Bessel functions 
of order $2L+1$. This can be seen by defining the variable
$x=\sqrt{|\beta_{LJ}|/r}$ and the function 
$\phi(x)=x^{-1/2}\psi_0(|\beta_{LJ}|/x^2)$, so that the equation becomes
\begin{equation}
\left[\frac{{\rm d}^2}{{\rm d}x^2}
+\frac{1}{x}\,\frac{{\rm d}}{{\rm d}x}-\frac{(2L+1)^2}{x^2}
\pm 4\right]\phi(x)=0,
\label{eq:zese2}
\end{equation}
where the plus sign applies to channels with even $J$ (where the 
tensor potential is attractive) and the minus sign to odd $J$.

The solutions in the attractive channels are oscillatory: 
\begin{equation}
\psi_0(r)=A\sqrt{\frac{|\beta_{LJ}|}{r}}\,\left[\sin\alpha\, 
J_{2L+1}\!\left(2\sqrt{\frac{|\beta_{LJ}|}{r}}\right)
+\cos\alpha\,
Y_{2L+1}\!\left(2\sqrt{\frac{|\beta_{LJ}|}{r}}\right)\right],
\label{eq:zeatt}
\end{equation}
where $J_{2L+1}$ and $Y_{2L+1}$  denote the regular and irregular Bessel 
functions. In the limit $r\rightarrow 0$, these solutions tend to the WKB 
form of a sinusoidal function of $2\sqrt{|\beta_{LJ}|/r}$ times $r^{-1/4}$.
They depend on a free parameter $\alpha$. This angle fixes the phase 
of the small-$r$ oscillations or, equivalently, it specifies a self-adjoint 
extension of the original Hamiltonian (see Refs.~\cite{bb2,birse} for 
further references). This is necessary since both Bessel functions
give acceptable solutions for small $r$. There is a redundancy between 
$\alpha$ and the leading term in the effective short-range potential since 
both have the effect of fixing the phase of the wave function for small $r$.
In waves where the scattering is weak, it is simplest to set $\alpha=0$
and use the potential to represent short-distance physics. This leads to 
solutions to Eq.~(\ref{eq:zese}) that grow like $r^L$ for large $r$. In 
channels where the OPE potential can be treated perturbatively, 
this allows the waves to match on to their usual short-distance forms at 
larger radii where the centrifugal barrier dominates over the tensor 
potential.

The special choice $\alpha=\pi/2$ gives a solution to Eq.~(\ref{eq:zese}) 
that decays like $r^{-(L+1)}$ for large $r$. Imposing this boundary 
condition on the full OPE problem would lead to a wave that was very large 
inside the attractive well of the $1/r^3$ potential. This 
would correspond to a system with a low-energy bound state or resonance. 
Since none of the NN channels with $L>0$ has such a low-energy state, 
values of $\alpha$ close to $\pi/2$ should be avoided. In practice, 
$\alpha=0$ is a good choice for all but one of the waves studied here. 

In the repulsive channels, the solutions are given by modified Bessel 
functions, and the regular one has the form
\begin{equation}
\psi_0(r)=A\sqrt{\frac{|\beta_{LJ}|}{r}}\,
K_{2L+1}\!\left(2\sqrt{\frac{|\beta_{LJ}|}{r}}\right).
\label{eq:zerep}
\end{equation}
This vanishes exponentially with $2\sqrt{|\beta_{LJ}|/r}$ as 
$r\rightarrow 0$ but, like Eq.~(\ref{eq:zeatt}), grows as $r^L$ for large $r$. 

It is convenient to normalise these solutions so that, as $r$ increases, 
they match on to the expected short-distance behaviour of the free solutions, 
$j_L(pr)/p^L$. (Since the asymptotic solutions are defined at zero energy,
I have divided out the $p^L$ energy dependence from the spherical Bessel 
functions.) With this normalisation, 
\begin{equation}
\psi_0(r)\sim \frac{r^L}{(2L+1)!!},\qquad\mbox{as}\ r\rightarrow\infty.
\label{eq:sdsphbes}
\end{equation}
the asymptotic solutions become
\begin{equation}
\psi_0(r)=\left\{\begin{array}{ll}
-\,{\displaystyle\frac{\pi|\beta_{LJ}|^L}{(2L)!(2L+1)!!\cos\alpha}}\,
\sqrt{{\displaystyle\frac{|\beta_{LJ}|}{r}}}\,\left[\sin\alpha\, 
J_{2L+1}\!\left(2\sqrt{{\displaystyle\frac{|\beta_{LJ}|}{r}}}\right)
+\cos\alpha\,
Y_{2L+1}\!\left(2\sqrt{{\displaystyle\frac{|\beta_{LJ}|}{r}}}\right)\right]
&\quad J\ \mbox{even,}\cr
\noalign{\vspace{5pt}}
{\displaystyle\frac{2|\beta_{LJ}|^L}{(2L)!(2L+1)!!}}\,
\sqrt{{\displaystyle\frac{|\beta_{LJ}|}{r}}}\,
K_{2L+1}\!\left(2\sqrt{{\displaystyle\frac{|\beta_{LJ}|}{r}}}\right)
&\quad J\ \mbox{odd.}\end{array}\right.
\label{eq:zesolns}
\end{equation}

\begin{figure}[t,b]
\includegraphics[width=17.5cm,keepaspectratio,clip]{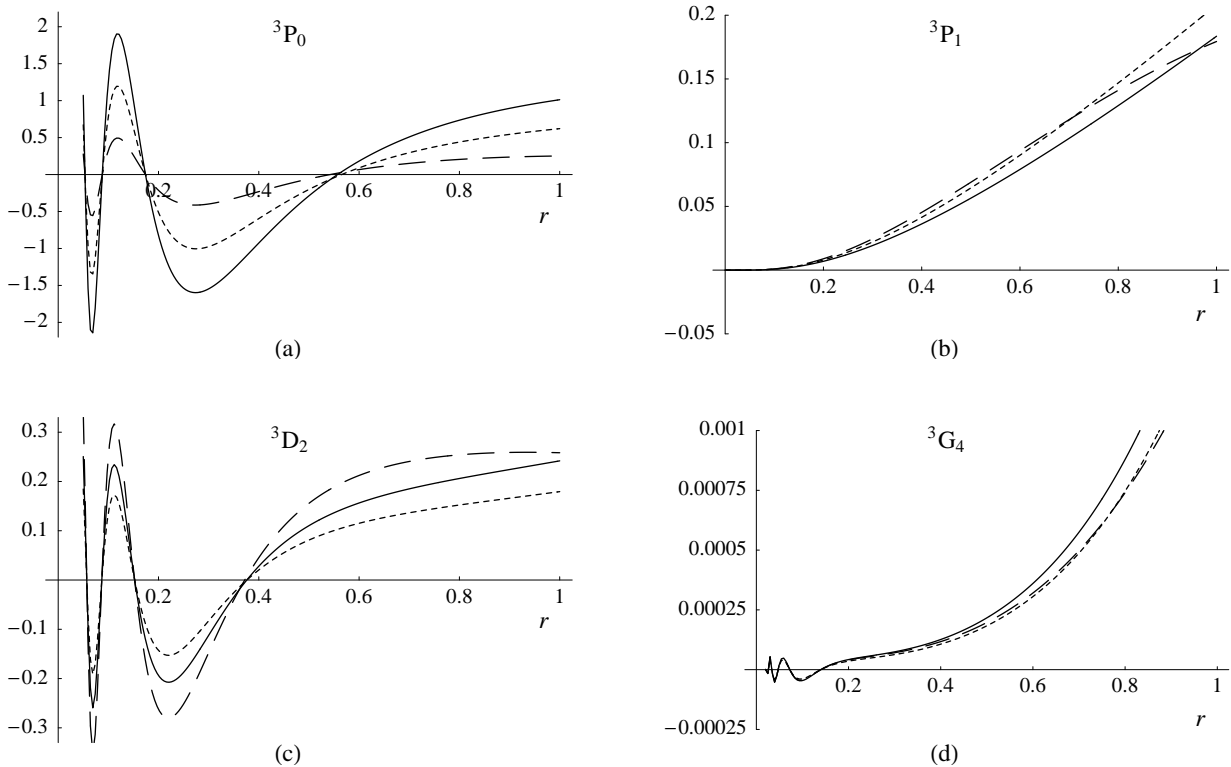}
\caption{\label{fig:wfns}  Plots of the wave functions $\psi(r)/p^L$ (in 
arbitrary units) against $r$ (in fm) for the channels (a) $^3P_0$, (b) $^3P_1$,
(c) $^3D_2$, and (d) $^3G_4$. Short-dashed lines: $T=5$~MeV; 
long-dashed lines: $T=300$~MeV; solid lines: energy-independent 
asymptotic form from Eq.~(\ref{eq:zese}).}
\end{figure}

Fig.~\ref{fig:wfns} shows the solutions to the full Schr\"odinger equation
(\ref{eq:se}) for the channels $^3P_0$, $^3P_1$, $^3D_2$ and $^3G_4$ at 
lab kinetic energies of 5 and 300~MeV, and compares them to the 
zero-energy, chiral-limit solutions of Eq.~(\ref{eq:zesolns}). To make this
comparison easier, I have also divided out the $p^L$ energy dependence 
from the full solutions. The waves shown for the attractive channels all 
have the short-distance phase $\alpha=0$. 

Although there is still some energy dependence in their normalisations,
the shapes of the solutions in these channels reach their energy-independent 
forms for radii smaller than about $0.8$~fm. The waves in channels where the 
tensor potential is attractive all oscillate for small enough radii. 
In the lowest waves, $^3P_0$ and $^3D_2$, this behaviour covers the whole
energy-independent region. Indeed the first node of the $^3P_0$ wave 
function appears at the edge of or, depending on the choice of $\alpha$, 
within the domain of the low-energy EFT. Moreover the normalisation of the 
short-distance wave functions shows significant energy-dependence, beyond 
the $p^L$ expected from the free solutions. 
 
In contrast, the $^3G_4$ wave becomes oscillatory only for radii smaller than 
about 0.25~fm, beyond the scope of our EFT. In the rest of its energy-independent 
region it has power-law behaviour as expected from the centrifugal potential.
More importantly from the EFT point of view, the energy dependence of the 
short-distance normalisation is almost entirely given by the $p^L$ factor
expected for a free solution. These results are consistent with the 
estimates in Ref.~\cite{birse} based on the chiral limit of the tensor 
potential. There, the critical momentum scales for the breakdown of 
perturbation theory were found to be of the order of $m_\pi$ or 
$\lambda_\pi$ in the $S$, $P$ and $D$ waves, and much larger ($\sim 2$~GeV) 
in the $F$ waves and above.

The $^3D_2$ and $^3G_4$ waves have low-energy bound states or narrow 
resonances if $\alpha$ is taken to be close to $\pi/2$. The
phase shifts can depend significantly on the choice of $\alpha$ in 
this region which, in the $^3D_2$ case, is roughly 
$\pi/4\lesssim\alpha\lesssim 3\pi/4$. The high centrifugal barrier in 
the $^3G_4$ wave means its phase shift is very weakly dependent on $\alpha$, 
outside a very narrow band around $\pi/2$. In contrast the $^3P_0$ phase 
shift shows a strong dependence on $\alpha$ over the whole range
$0\leq\alpha<\pi$.

The wave functions in the repulsive channels $^3P_1$ and $^3F_3$ both have 
qualitatively similar forms, and so only one is shown in Fig.~\ref{fig:wfns}. 
They switch from power-law behaviour at larger radii, where the 
centrifugal barrier dominates, to exponential at small radii, where the 
$1/r^3$ potential wins out. In both cases the energy dependence of the 
short-distance normalisation is quite well described by the free $p^L$ form, 
at least for energies below about 300 MeV. At higher energies than 
considered here, 500 MeV or above, the $^3P_1$ normalisation does develop 
a much stronger energy dependence (whereas the $^3F_3$ does not). These
numerical observations indicate that the effect of the finite 
pion mass has been to shift the critical momentum scale in $^3P_1$ channel
to a somewhat higher value than the estimate in Ref.~\cite{birse}.
From the wave functions alone, it is thus not clear whether tensor
OPE is better treated nonperturbatively in this channel.

\section{One-pion-exchange effective potential}

The RG method developed in Refs.~\cite{bb1} shows that the terms in
the short-range effective potential are directly related to an expansion of 
the DW $K$-matrix in powers of the energy and other low-energy scales.
A DW approach like this is obviously needed in channels where OPE must
be treated nonperturbatively. I will also use it in the channels where a 
perturbative treatment of OPE would be valid, since it provides a convenient 
alternative to fourth-order perturbation theory. The scattering between DW's 
of the long-range potential can be described by a reactance matrix 
$\widetilde K$. In the uncoupled channels, its on-shell matrix element 
is related to the difference between the observed and OPE phase shifts by 
\begin{equation}
\widetilde K(p)=-\,\frac{4\pi}{Mp}\,
\tan\Bigl(\delta_{\scriptscriptstyle\rm PWA}(p)
-\delta_{\scriptscriptstyle\rm OPE}(p)\Bigr),
\end{equation}
where $\delta_{\scriptscriptstyle\rm PWA}(p)$ is the empirical phase shift 
(from the Nijmegen group's PWA or one of their potentials that have
been fitted directly to data) and
$\delta_{\scriptscriptstyle\rm OPE}(p)$ is that obtained from
the solutions of Eq.~(\ref{eq:se}). 
None of the channels examined here contains a low-energy bound state or 
resonance, and so the residual scattering can be represented by an EFT 
expanded around a trivial fixed point. The amplitude $\widetilde K(p)$ is 
then given by the distorted-wave Born approximation (DWBA): the matrix 
element of the short-range interaction between the DW's of the long-range 
potential.

Since these DW's are either vanishing or singular as $r\rightarrow 0$, an 
ordinary $\delta$-function at the origin cannot be used to represent the 
potential. Instead, I take a $\delta$-shell potential, as in 
Refs.~\cite{bb1,bmcg,birse}. If the radius of this, $R_0$, is chosen to 
be smaller than about 0.7~fm, in the region where the waves have reached 
their common energy-independent forms, then the extracted potential 
will be independent of $R_0$ to a good approximation.\footnote{Obviously, 
points where the wave functions vanish in the attractive channels should be 
avoided for numerical reasons. Similarly very small values of $R_0$, 
$\sim 0.01$~fm or less, should not be used in the repulsive channels.}
The results shown below are for $R_0=0.1$~fm, the same radial cut-off
as used in Ref.~\cite{bmcg}, but I return to the question of the choice of 
cut-off at the end of Sec.~IV, after examining the subtraction of TPE.

To remove the dependence on the arbitrary radius $R_0$, 
I divide out the square of the asymptotic radial function, 
Eqs.~(\ref{eq:zeatt}) or (\ref{eq:zerep}), from the strength of the 
$\delta$-shell potential. The resulting potential is defined by
\begin{equation}\label{eq:vshort}
V_S(p,r)=\frac{1}{4\pi R_0^2|\psi_0(R_0)|^2}\,\widetilde V(p)\,\delta(r-R_0),
\end{equation}
as in the RG analysis of Refs.~\cite{bb1,birse}.
Equating $\widetilde K(p)$ to the DWBA matrix element of this potential,
we can deduce the strength of the potential directly from the residual 
scattering amplitude, as in Ref.~\cite{bmcg}:
\begin{equation}\label{eq:vtilde}
\widetilde V(p)=\frac{|\psi_0(R_0)|^2}{|\psi(p,R_0)|^2}\,\widetilde K(p),
\end{equation}
where $\psi(p,R_0)$ is the solution to the Schr\"odinger equation
with the known long-range potential, and $\psi_0(R_0)$ is its
energy-independent short-distance form.

The leading term in the short-distance interaction represents only
short-distance physics, and so it should be independent of any low-energy
(long-distance) scales such as $p$, $m_\pi$ or $\lambda_\pi$. To ensure this,
any dependence on these scales should be factored out of the normalisation 
of the asymptotic solutions in Eqs.~(\ref{eq:zeatt}) and (\ref{eq:zerep}). 
In waves where the OPE potential can be treated perturbatively, this means 
normalising these functions by dividing out the energy-dependent factor of 
$p^L$, as in Eq.~(\ref{eq:zesolns}). The resulting potentials in the 
$^3F_3$ and $^3G_4$ channels are the defined in the same way as the ones 
used to analyse the spin-singlet channels in Ref.~\cite{bmcg}. 
At LO their matrix elements are proportional to $p^{2L}$,
showing that they are equivalent to $2L$-th derivatives of  
$\delta$-functions (the more conventional representations for 
the effective interactions in these partial waves).

As discussed above, $\lambda_\pi$ should be regarded as an additional 
low-energy scale in the channels where OPE must be treated 
nonperturbatively. Since $\beta_{LJ}\propto 1/\lambda_\pi$, the wave 
functions normalised as in Eq.~(\ref{eq:zesolns}) still contain powers of 
$\lambda_\pi$. This can be removed by dividing out the factors of 
$|\beta_{LJ}|^{L+1/4}$ from the zero-energy solutions. Using solutions 
with this normalisation has the effect of multiplying $\widetilde V(p)$ by 
$\lambda_\pi^{2L+1/2}$. This obviously changes the magnitudes of the 
short-range interactions, but it does not affect their energy 
dependences. In the plots below, I just show results for $\widetilde V(p)$ 
defined using the forms given in Eq.~(\ref{eq:zesolns}), but the extra 
factors would be needed if one wanted to estimate the momentum scales
in this potential.

Once the effects of leading OPE have been removed, the residual potential 
starts at order $Q^2$ in the standard power-counting notation.
Contributions at this order include the leading TPE potential 
\cite{kbw,nij99}, and a relativistic correction to OPE \cite{friar}. 
Since the power counting for the long-range forces is not affected by
the renormalisation of the short-range ones, I denote this potential
by $\widetilde V^{(2)}(p)$. In Weinberg's counting for a partial wave 
with angular momentum $L$, the leading short-range term appears at the 
usual order, $Q^{2L}$. However, if OPE is treated nonperturbatively, 
we need to take account of the fact that this term also contains
a factor of $\lambda_\pi^{-(2L+1/2)}$. Since $\lambda_\pi$ 
should be now regarded as a low-energy scale, we see that the 
net order of such a term is $Q^{-1/2}$, for any $L$. This half-integer 
power agrees with the RG analysis in Ref.~\cite{birse}, which shows 
that the leading term has an RG eigenvalue of $1/2$. After extracting 
the effects of tensor OPE we are therefore left with an interaction 
that starts at order $Q^{-1/2}$. This also contains a term proportional 
to energy ($p^2$) at order $Q^{3/2}$. 

\begin{figure}[h]
\includegraphics[width=17.5cm,keepaspectratio,clip]{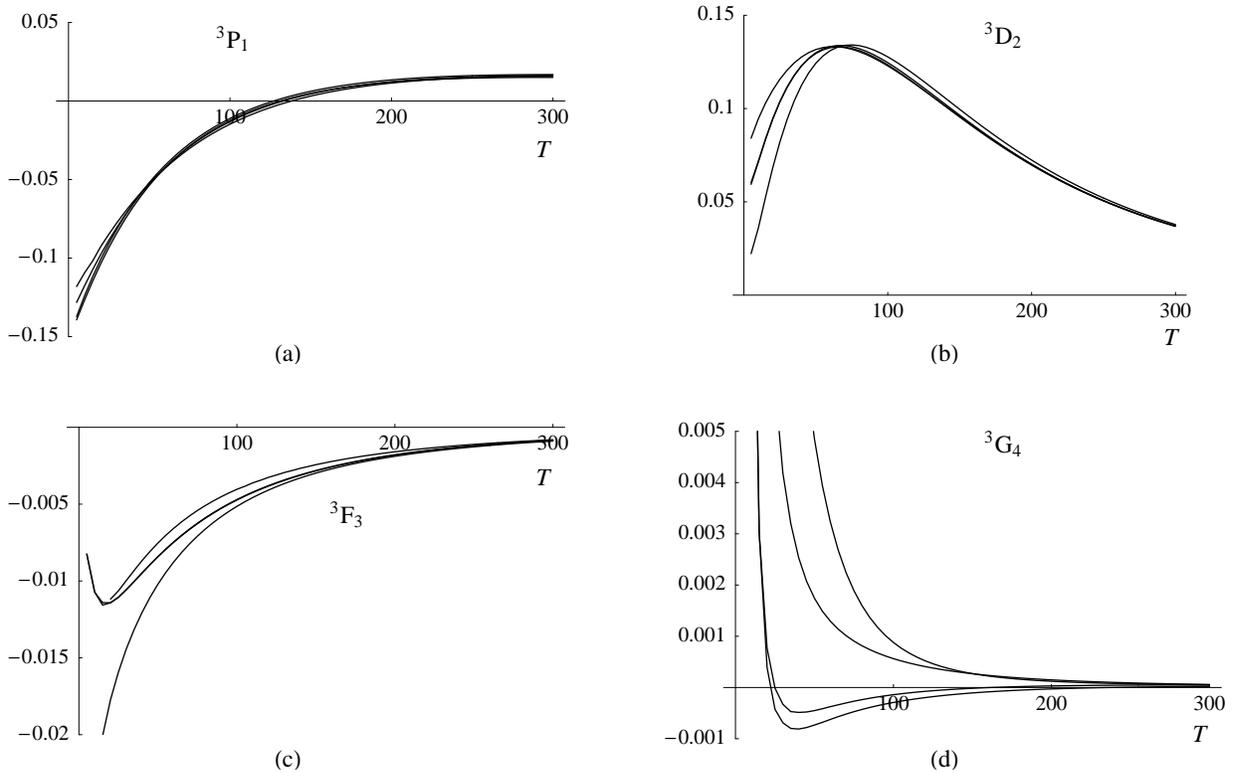}
\caption{\label{fig:ps1234}  Plots of the short-distance interaction 
$\widetilde V^{(2)}(p)$, in fm$^{2L+2}$, against lab kinetic energy $T$, 
in MeV. These have been extracted from Nijmegen PWA's or potentials 
using Eq.(\ref{eq:vtilde}) with $R_0=0.1$~fm, for the $np$ channels 
(a) $^3P_1$, (b) $^3D_2$, (c) $^3F_3$, and (d) $^3G_4$.}
\end{figure}

Fig.~\ref{fig:ps1234} shows the short-distance interactions 
$\widetilde V^{(2)}(p)$ extracted directly from various Nijmegen 
analyses \cite{nijnn}, using Eq.~(\ref{eq:vtilde}) with $R_0=0.1$~fm. 
The partial-wave analysis, PWA93, and three potentials, NijmegenI, 
NijmegenII and Reid93, all fit the $np$ data with similarly good values 
of $\chi^2$. They can thus be regarded as alternative partial-wave 
analyses. Using results from all of them gives an indication of the
systematic uncertainties associated with the different choices of 
parametrisation. This is particularly important in higher partial waves 
where each fit contains only a small number of parameters. In the 
$^3D_2$ and $^3G_4$ channels where the tensor OPE is attractive, I have 
taken the phase $\alpha=0$ since, as already noted, their OPE phase shifts 
depend only weakly on $\alpha$, provided the region around $\pi/2$ is 
avoided.

One immediate observation is that the residual interactions all show
rapid, nonlinear dependences on energy below about 150~MeV. This is 
similar to what was found for the corresponding spin-singlet channels
in Ref.~\cite{bmcg}. Experience with those channels suggests that TPE 
is responsible for much of this rapid energy dependence. Therefore
these forces need to be subtracted before any conclusions can be drawn 
about short-range ones.

\begin{figure}[h]
\includegraphics[width=17.5cm,keepaspectratio,clip]{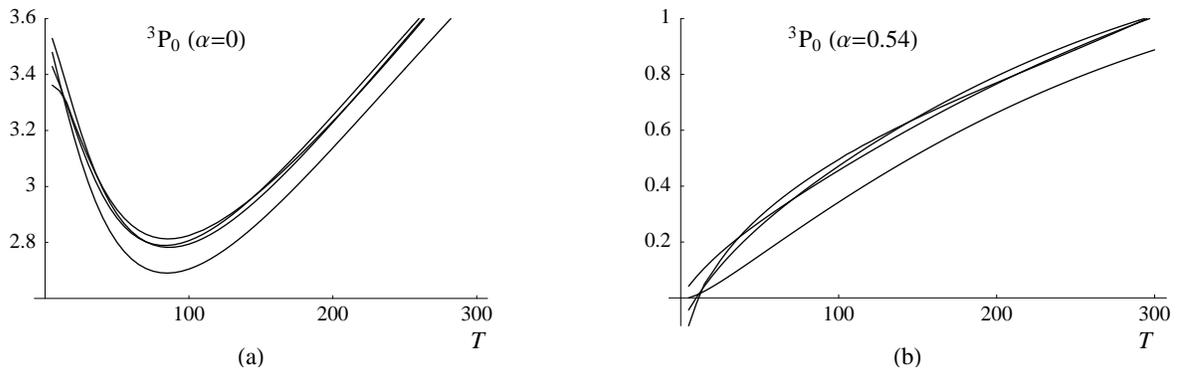}
\caption{\label{fig:ps3p0}  Plots of the $^3P_0$ short-distance 
interaction $\widetilde V^{(2)}(p)$, in fm$^{4}$, against lab kinetic 
energy $T$, in MeV. Results for two choices of short-distance
phase are shown: (a) $\alpha=0$, (b) $\alpha=0.54$.}
\end{figure}

Finally I turn to the $^3P_0$ wave, shown in Fig.~\ref{fig:ps3p0}. 
The left-hand plot was obtained using $\alpha=0$, as in the $^3D_2$ 
and $^3G_4$ cases. As in the attractive $^3D_2$ channel, this shows 
a rapid, non-monotonic energy dependence.
In addition, its overall strength is much larger than in the $^3P_1$
channel, by a factor of 30 or more. This implies that the leading 
(order-$Q^{-1/2}$) interaction in this channel is too strong to be
treated in the DWBA. It should either be iterated to higher 
orders or, possibly, treated nonperturbatively.

It practice, it is most convenient to include this term to all
orders using the equivalent short-distance parameter $\alpha$.
As discussed above, this defines a self-adjoint extension of the
the OPE Hamiltonian in Eq.~(\ref{eq:se}) by fixing the phase of 
the oscillations of the wave functions as $r\rightarrow 0$.
In the right-hand plot, I show the residual interaction
for $\alpha=0.54$. This value was chosen so that the 
interaction vanishes at zero energy, within the uncertainties of the
PWA's. Removing the effects of the energy-independent term
in this way leaves a residual interaction that is consistent with
a monotonic piece plus one linear in the energy.
Again, more definite conclusions require subtraction of the effects
of other long-range forces, which I now turn to.

\section{Two-pion exchange}

The DW method described above could also be used to separate
off the scattering produced by other known long-range forces, 
most importantly those arising from two-pion exchange. However, 
since such forces start at order $Q^2$, their effects can just 
be subtracted perturbatively from the residual interactions
left after removal of OPE \cite{bmcg}. Although iterations of TPE 
will contribute at higher orders, this DWBA treatment is adequate
up to order $Q^4$.

The relevant TPE potentials have been calculated at orders $Q^2$ and 
$Q^3$ and their forms can be found in Refs.~\cite{kbw,nij99}.
At order $Q^2$, the long-range interactions also include a term
from expanding the relativistic correction to OPE \cite{friar},
\begin{equation}\label{eq:ope2}
V^{(2)}_{1\pi}(r)=-\frac{p^2}{2M^2}\left[V_{\pi {\scriptscriptstyle C}}(r)
+V_{\pi {\scriptscriptstyle T}}(r)\right].
\end{equation}
Lastly, there is an electromagnetic interaction, generated by $\pi\gamma$ 
exchange, whose form is derived in \cite{fvkpc}.

If all of these are subtracted from $\widetilde K(p)$ using the DWBA, then
any residual long-ranged contributions to the scattering start at order
$Q^4$. The resulting potential is
\begin{equation}\label{eq:vt4}
\widetilde V^{(4)}(p)=\frac{|\psi_0(R_0)|^2}{|\psi(p,R_0)|^2}
\left(\widetilde K(p)-\langle\psi(p)|V_{1\pi}^{(2)}
+V_{2\pi}^{(2,3)}+V_{\pi\gamma}|\psi(p)\rangle\right).
\end{equation}
Note that although this potential is denoted $\widetilde V^{(4)}(p)$
because it contains long-range contributions of order $Q^4$ and higher, 
it may contain short-range terms of lower order. Such terms appear
first at order $Q^2$ for $P$-waves in the usual power counting, 
and at order $Q^{-1/2}$ for waves where tensor OPE is treated 
nonperturbatively.

The long-range potentials are singular and so their matrix elements 
can be divergent. I therefore cut the integrals off, using the same radius, 
$R_0$, as in the definition of the short-range interaction. For example, 
the most divergent part of the order-$Q^3$ TPE potential at small radii 
is proportional to $1/r^6$ \cite{kbw,nij99}. In a perturbative treatment 
of OPE, the wave functions behave like $r^L$ as $r\rightarrow 0$. Together 
these lead to a $1/R_0$ divergence in the $P$-wave matrix elements 
which can be renormalised by introducing the order-$Q^2$ counterterm 
just mentioned. Other partial waves do not give rise to divergences to 
this order.

When OPE is treated nonperturbatively the small-$r$ forms of the wave 
functions are quite different, as described above. The $r^{-1/4}$
behaviour of these leads to a $1/R_0^{7/2}$ divergence in
the matrix element of the order-$Q^3$ potential, at
least in the channels where tensor OPE is attractive. This can be
renormalised by the leading short-distance potential, 
of order $Q^{-1/2}$. In addition, there are other, weaker
divergences which appear multiplied by powers of $\lambda_\pi$ or 
$m_\pi$ and which can be renormalised by higher-order counterterms.
However these energy-independent terms can not be disentangled 
phenomenologically from the leading one. The relativistic
correction to OPE, Eq.~(\ref{eq:ope2}), also leads to a  
divergence and this can be renormalised by an order-$Q^{3/2}$ term, 
proportional to $p^2$. The complete set of terms with orders
below $Q^4$ also includes one proportional to the square of the 
energy, $p^4$. This term is of order $Q^{7/2}$ and it has a finite 
coefficient at the current level of approximation. However, when 
long-range forces of order $Q^4$ are included, it will be needed to 
renormalise divergences from, for example, the next relativistic 
correction to OPE.

The short-range terms up to order $Q^{7/2}$ provide a smooth, 
quadratic energy-dependence. Subtracting them
should leave a residual interaction consisting only of terms of
order $Q^4$ or higher. I do this by fitting a quadratic form,
\begin{equation}
\widetilde V^{(7/2)}(p)=C_0+C_2p^2+C_4p^4,
\end{equation}
to $\widetilde V^{(4)}(p)$ in the range $T=100$ to 200~MeV.
The lower end of this range is chosen to lie above the region where OPE 
and TPE can lead to rapid energy dependence. The upper limit corresponds
to a relative momentum $p\simeq 300$~MeV. Above this point, higher
powers of the the energy could start to become noticeable, representing
physics that has been integrated out, such as excitation of the $\Delta$
resonance.

\begin{figure}[h]
\includegraphics[width=17.5cm,keepaspectratio,clip]{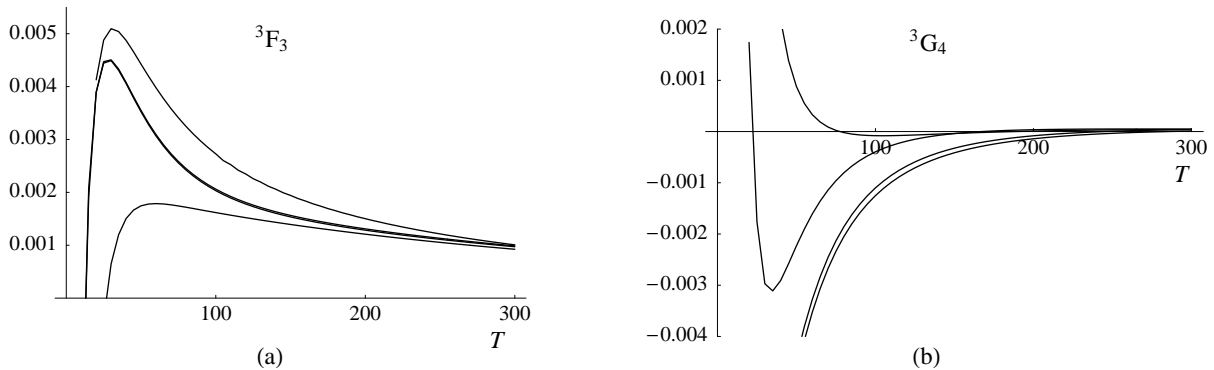}
\caption{\label{fig:v34}  Plots of the short-distance 
interaction $\widetilde V^{(4)}(p)$, in fm$^{2L+2}$, against lab kinetic 
energy $T$, in MeV, for the channels: (a) $^3F_3$, (b) $^3G_4$.}
\end{figure}

Let me start with the two peripheral waves, where the breakdown scale 
for perturbation theory is so high that there is no question about 
the validity of the perturbative treatment of OPE or the standard 
power counting. As in Ref.~\cite{bmcg}, I use the full DW solutions
for these, although these are not strictly necessary, to avoid the
complications of fourth-order perturbation theory. The residual 
interactions $\widetilde V^{(4)}(p)$ for these are shown in 
Fig.~\ref{fig:v34}. In both cases, $\widetilde V^{(4)}(p)$ was extracted at 
$R_0=0.1$~fm, which is small enough that the wave functions have attained 
their energy-independent forms and that the $^3F_3$ radial integral has 
converged. The $^3G_4$ integral has reached a plateau here; the divergences
associated with the oscillatory region do not appear until $R_0$ is less 
than about 0.03~fm.

As in the corresponding spin-singlet waves \cite{bmcg}, the uncertainties 
in the PWA's make it hard to draw strong conclusions. The $^3G_4$ residual 
interactions, $\widetilde V^{(2)}(p)$ and $\widetilde V^{(4)}(p)$, are both 
consistent with zero, given the spread of the PWA's. In the $^3F_3$ 
channel, the energy dependence in $\widetilde V^{(2)}(p)$ below 
100~MeV is somewhat reduced by subtraction of the order-$Q^{2,3}$ 
long-range forces. 

\begin{figure}[h]
\includegraphics[width=17.5cm,keepaspectratio,clip]{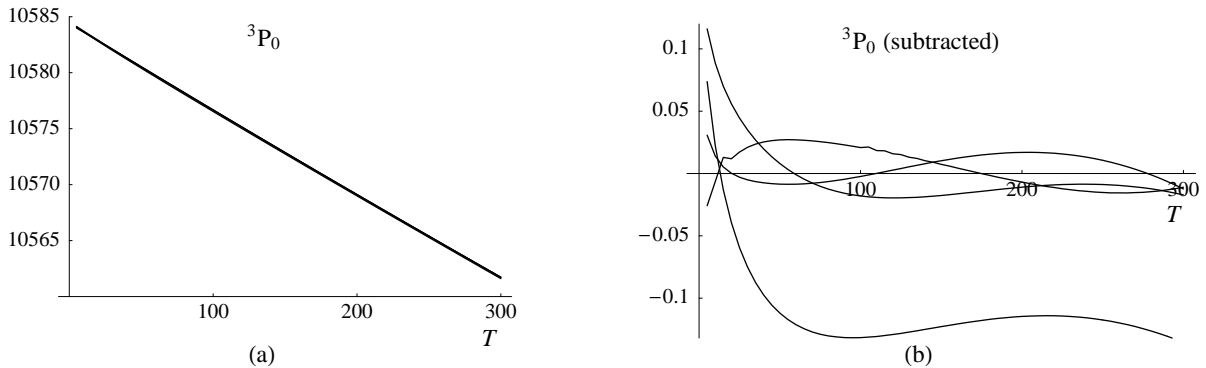}
\caption{\label{fig:v0}  Plots of the short-distance 
interaction $\widetilde V^{(4)}(p)$, in fm$^{4}$, against lab kinetic 
energy $T$, in MeV, for the channel $^3P_0$ with $\alpha=0.54$, (a) 
unsubtracted, (b) quadratic fit to $T=100-200$~MeV subtracted.}
\end{figure}

The interaction $\widetilde V^{(4)}(p)$ in the $^3P_0$ channel is shown in 
Fig.~\ref{fig:v0}. The unsubtracted interaction, on the left, is very large, 
with a strong linear energy dependence, as a result of the divergences in the 
integrals of the order-$Q^{2,3}$ potentials. The residual interaction after 
subtracting a quadratic fit is shown on the right.\footnote{In this channel, I 
have omitted the Reid93 results from the fit since they lie systematically 
below the ones from the other three Nijmegen analyses. Including this potential 
would simply shift the fitted constant, and significantly increase the 
uncertainty associated with the PWA's.} This shows that the quadratic 
form can provide a good account of the energy dependence of 
$\widetilde V^{(4)}(p)$ in the $^3P_0$ channel over the whole range of $T$ 
up to 300~MeV. Any contributions from forces of order $Q^4$ and higher lie 
within the uncertainties of the PWA's.

\begin{figure}[h]
\includegraphics[width=17.5cm,keepaspectratio,clip]{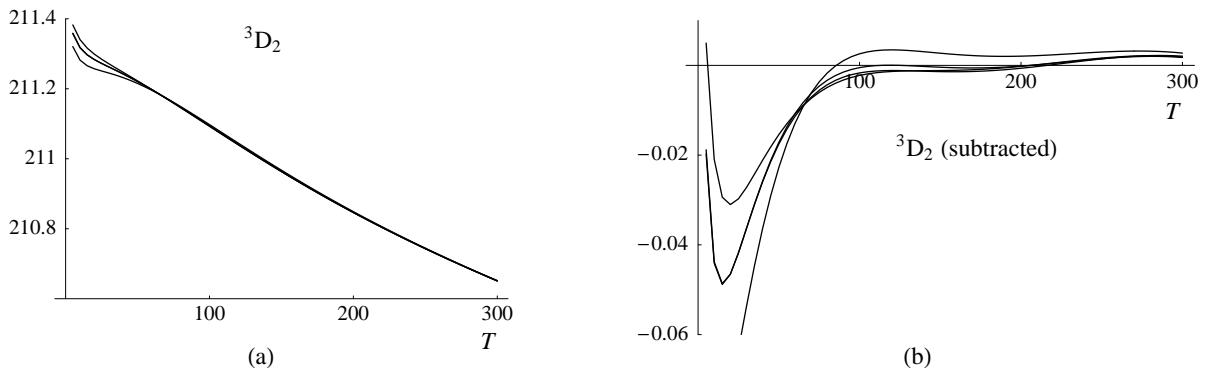}
\caption{\label{fig:v2}  Plots of the short-distance 
interaction $\widetilde V^{(4)}(p)$, in fm$^{6}$, against lab kinetic 
energy $T$, in MeV, for the channel $^3D_2$ with $\alpha=0$, (a) 
unsubtracted, (b) quadratic fit to $T=100-200$~MeV subtracted.}
\end{figure}

Fig.~\ref{fig:v2} shows the results of a similar analysis of scattering
in the $^3D_2$ channel. Here, the quadratic fit can account well for the 
energy dependence from $T\simeq 80$~MeV to about 250~MeV. Below 80~MeV, 
the PWA's require a significant extra attractive interaction with a long 
range in view of its rapid energy dependence. It is not obvious what could 
be responsible for this, since no appropriate piece seems to 
be present in the order-$Q^4$ pion-exchange potentials \cite{kaiser,machl}. 
However it should be noted that its size is comparable to the uncertainties 
in the PWA's in this region.

\begin{figure}[h]
\includegraphics[width=17.5cm,keepaspectratio,clip]{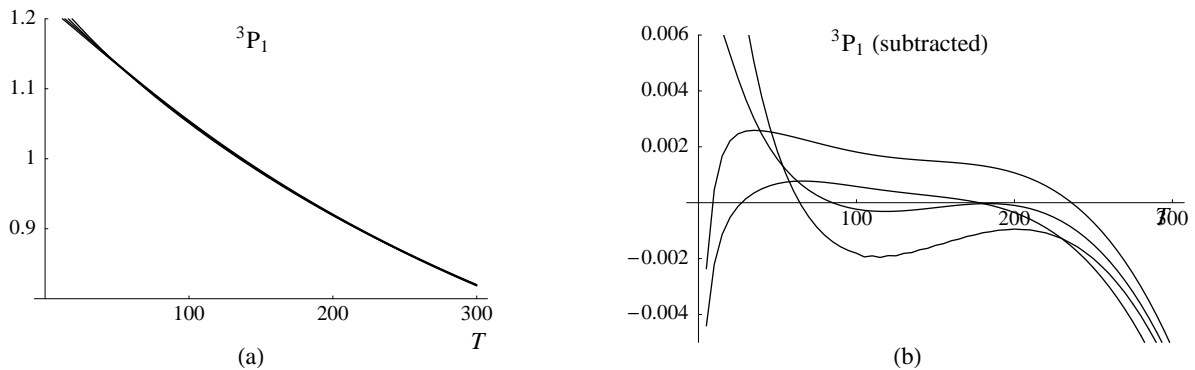}
\caption{\label{fig:v1}  Plots of the short-distance 
interaction $\widetilde V^{(4)}(p)$, in fm$^{4}$, against lab kinetic 
energy $T$, in MeV, for the channel $^3P_1$, (a) unsubtracted, 
(b) quadratic fit to $T=100-200$~MeV subtracted.}
\end{figure}

Finally, $\widetilde V^{(4)}(p)$ for the $^3P_1$ channel has a smooth, 
approximately linear energy dependence, as can be seen in Fig.~\ref{fig:v1}. 
The value at zero energy increases as the cut-off radius $R_0$ is decreased 
from 0.6~fm to about 0.2~fm. This is consistent with the standard power 
counting, where a divergence is present at order $Q^2$. In that counting, 
this is the only divergence below order $Q^4$. However the coefficient of 
the linear energy dependence in the results here also increases significantly 
over this range of $R_0$. This suggests that the $^3P_1$ channel may in fact
be better analysed using the new power counting, a conclusion that is
supported by the low breakdown scale for the perturbative treatment of OPE
found in Ref.~\cite{birse}.

The results shown so far are all obtained with a radial cut-off
$R_0=0.1$~fm. This is well inside the region where the dependence on 
$R_0$ is very small for the range of energies considered but it does 
correspond to momentum scales greater than 1~GeV, 
far beyond the range of validity of our EFT. This small radial cut-off
was used in Ref.~\cite{bmcg} to avoid introducing artefacts proportional 
to positive powers of $R_0$. The coefficients in the residual potential 
could then be related directly to scales of the underlying physics. The
connection is more complicated here, as one would first need to renormalise 
the coefficients to remove the divergences proportional to powers of 
$1/R_0$ (and there are several of these in the energy-independent term, 
involving different powers of $m_\pi$ and $\lambda_\pi$.) As a result,
I do not attempt to make such an interpretation of the coefficients here
and hence the choice of such a small radius in not crucial.

In this case, one can ask what happens if a larger cut-off radius 
is chosen. The question is pertinent since Epelbaum and Meissner 
\cite{em} have shown that a single extra term, as required by the new
counting, can give a fairly good description of $P$- and $D$-wave phase 
shifts for momentum cut-offs as low as $\sim 3$~fm$^{-1}$. Such values 
would also avoid regions where TPE might need to be treated 
nonperturbatively. I have therefore repeated these analyses with larger 
cut-off radii. The results are essentially indistinguishable from the 
ones shown above for radii up to about 0.6~fm (except in regions around 
zeros of the wave functions). Beyond that point, the wave functions do 
show noticeable energy dependence over the range up to $T=300$~MeV, as 
already noted from Fig.~\ref{fig:wfns}. This leads to a dependence on 
$R_0$ of the residual interactions, especially at higher energies. 

If these artefacts of a finite radial cut-off are expanded as a power 
series in $R_0^2p^2$, the first three terms (up to order $p^4$) can be 
absorbed in the coefficients of the quadratic fit. They are therefore 
removed when this fitted potential is subtracted as described above. 
However, they will generate terms in the renormalised coefficients of 
the short-range potential where the momentum scale is set by $1/R_0$ 
and not by the underlying physics. This would make it difficult to 
interpret these coefficients in terms of physical scales.

This cut-off dependence increases with energy and so it is most 
prominent at higher energies. A measure of how much the shape of the 
waves changes is provided by the ratio of ratios,
\begin{equation}
\rho=\left|\frac{\psi(p_{\rm max},R_0)\psi(0,R_1)}
{\psi(p_{\rm max},R_1)\psi(0,R_0)}\right|^2,
\end{equation}
in which the energy-dependent normalisation of the short distance 
wave functions cancels out. For $R_0=0.6$~fm, $R_1=0.1$~fm and 
$p_{\rm max}=375$~MeV (corresponding to $T=300$~MeV), $\rho$ is 
greater than 0.8 for most of the waves considered here. The $\sim 20\%$ 
changes that this induces in the the residual interactions can 
be removed by the subtraction of the quadratic fit. The resulting 
subtracted interactions are almost indistinguishable from the ones 
shown in Figs.~\ref{fig:v2} (b) and \ref{fig:v1} (b), 
except for the $^3P_0$ which shows larger effects as a result of
a nearby zero in its wave functions. 

\begin{figure}[h]
\includegraphics[width=17.5cm,keepaspectratio,clip]{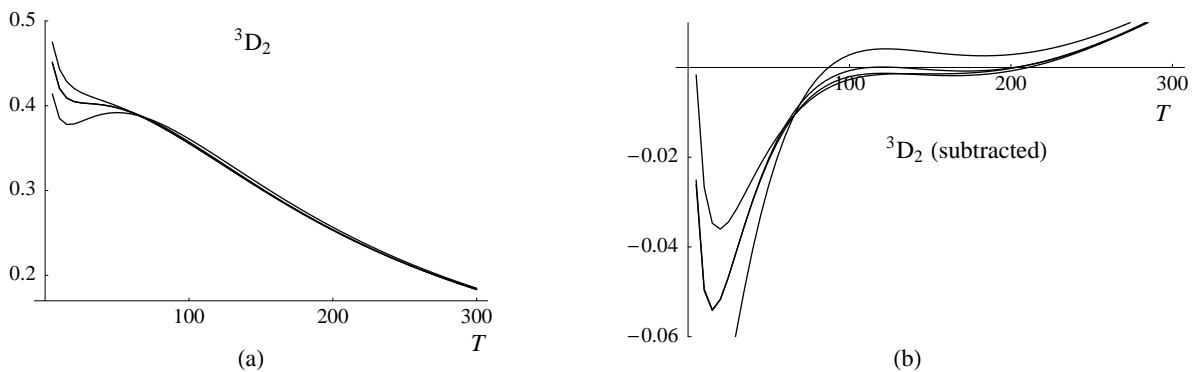}
\caption{\label{fig:v2n}  Plots of the short-distance 
interaction $\widetilde V^{(4)}(p)$ for the channel $^3D_2$ with cut-off 
radius of 1.0~fm. Other details are the same as in Fig.~\ref{fig:v2}.}
\end{figure}

For $R_0=1.0$~fm, the ratio $\rho$ is around 0.6 for these waves.
The residual interaction $\widetilde V^{(4)}(p)$ for the $^3D_2$ 
channel is shown in Fig.~\ref{fig:v2n}. The divergent terms are much 
smaller and so the overall magnitude of the potential is greatly 
reduced compared to the one shown in Fig.~\ref{fig:v2} (a). However 
the bulk of the change lies in the terms up to order $p^4$. After 
these have been subtracted, the differences between Figs.~\ref{fig:v2} 
(b) and \ref{fig:v2n} (b) are small, at least for energies below about 
250~MeV. The pattern in the $^3P_1$ channel is similar. The $^3P_0$ is 
complicated by the fact that the wave functions pass through zero near 
0.8~fm. Cut-off-independent results for the subtracted interaction can 
be obtained, but only for $R_0\lesssim 0.5$~fm or $R_0\simeq 1.2$~fm.
In the higher partial waves, $^3F_3$ and $^3G_4$, the unsubtracted 
residual interactions show very little cut-off dependence over the
range $0.02\ {\rm fm}\lesssim R_0\lesssim 1$~fm.

\section{Conclusions}

Nogga, Timmermans and van Kolck \cite{ntvk} have found that a new 
power counting is needed to organise the EFT describing
nucleon-nucleon scattering in spin-triplet channels. In particular,
the leading short-distance terms in $P$ and $D$ waves are significantly 
promoted compared to the perturbative (Weinberg) counting. In 
Ref.~\cite{birse} I obtained a quantitative statement of this new 
counting by identifying the scale of the OPE potential, $\lambda_\pi$, 
as an additional low-energy scale, and then treating OPE nonperturbatively.
This leads to an expansion of the short-range interaction 
describing scattering between the DW's of the OPE potential. 
Its terms correspond to those of an expansion of the DW Born 
amplitude in powers of the energy. The leading term is of
order $Q^{-1/2}$, for any orbital angular momentum.

The forms of the DW's show that a nonperturbative treatment of tensor OPE, 
and hence the new power counting, is required for energies that are large 
enough for the waves to probe the region where the $1/r^3$ core of the 
potential dominates over the centrifugal barrier. Otherwise the 
short-distance wave functions have the normal $r^L$ behaviour and 
perturbative counting remains valid. The analytic estimates 
in Ref.~\cite{birse} and the numerical wave functions in Sec.~II both 
indicate that the nonperturbative approach is required in waves with 
$L\leq 2$ for momenta of the order of $m_\pi$ or larger. In $F$ waves and 
above, a perturbative treatment is expected to remain valid 
up to energies well beyond the validity of the EFT. 

Here I have ``deconstructed" empirical phase shifts by using DW methods to 
remove the effects of long-range pion-exchange forces. This generates a
residual short-range interaction directly from the observed phase shifts.
Unlike comparisons of phase-shift plots, this approach emphasises 
the low-energy region, where the EFT description ought to work 
best. The use of several Nijmegen partial wave analyses \cite{nijnn} 
allows estimates of the uncertainties in these fits to the data. These
can be large at low energies, implying that it may be misleading to
fit the coefficients of an EFT to very low-energy data.  

Removing only the effects of OPE leaves residual interactions with 
strong energy dependences at low energies. This suggests that 
higher-order long-range interactions are also important. After 
removing the effects of OPE, I therefore use the DWBA to subtract the 
contributions of other long-range potentials up to order $Q^3$.
These include TPE and a relativistic correction to OPE, all of which
highly singular at the origin. Their DWBA matrix elements diverge 
as inverse powers of the cut-off radius but these divergences can be 
cancelled using counterterms at the orders required by the new power 
counting. In contrast, Weinberg counting at order $Q^3$ would provide 
only one, energy-independent counterterm in each $P$ wave, and none in 
any higher wave. 

TPE and other long-range forces up to order $Q^3$ are able to account 
for much of rapid energy dependence seen below 100~MeV in the $P$ and 
$D$ waves. When these are subtracted the residual scattering
amplitudes can, in general, be well fitted by three contact terms 
up to order $Q^{7/2}$ in the new counting. The only exception is
the $^3D_2$ channel, which seems to require an additional long-range
attraction. It is not clear where this could arise, given the forms of
the order-$Q^4$ chiral potentials \cite{kaiser,machl}, but one should 
note that the uncertainties in the PWA's for this channel are 
significant below about 70~MeV. Otherwise, TPE and the short-range 
forces up to order $Q^{7/2}$ are able to give a good description of 
the scattering in these triplet waves up to energies of about 250~MeV.

In the more peripheral $F$ and $G$ waves, the arguments of 
Ref.~\cite{birse} suggest that the standard Weinberg counting should be 
adequate for the energies considered here. The numerical wave functions 
support this, their short-range forms having the expected $p^L$ 
dependence driven by the centrifugal barrier. Subtraction of the 
long-range forces leaves small residual interactions, as in the 
corresponding singlet channels studied in Ref.~\cite{bmcg}. Provided
the cut-off radius is larger than about 0.1~fm, there is no sign of 
any divergences whose renormalisation would require the new power 
counting.

These results indicate that the spin-triplet waves with 
$L\leq 2$ can be analysed consistently using the nonperturbative power 
counting developed in Refs.~\cite{ntvk,birse}. In addition, the results 
show that deconstructing scattering amplitudes, using the approach of 
Ref.~\cite{bmcg}, can provide a very useful tool for determining
effective potentials directly from empirical phase shifts. It should
be straightforward to extend the method to coupled channels, such as
$^3S_1$--$^3D_1$, despite the more complicated matrix equations involved.
It would also be interesting to use it to subtract two- and three-pion 
exchanges at order $Q^4$ \cite{kaiser,machl}. However the approach also 
demonstrates that there are significant uncertainties in the currently
available Nijmegen PWA's, particularly for energies below about 80 MeV
where the scattering is most sensitive to long-range forces. As a result,
attempts to study the importance of higher-order forces may require the 
newer PWA's of Refs.~\cite{nij99,nij03}, when these become available.

\section*{Acknowledgments}

I am grateful to E. Epelbaum, H. Griesshammer, J. McGovern, D. Phillips and 
U. van Kolck for helpful discussions.

\end{document}